\documentstyle[aps, prd]{revtex}
\begin{document}
\draft
\title{Parity Doublets in Quark Physics}
\author{A. P. Balachandran 
	 and S. Vaidya \footnote{sachin@suhep.phy.syr.edu}
}
\address{\em Department of Physics, Syracuse University,
	     \\ Syracuse, N. Y.  13244-1130,  U. S. A. 
	\\
}
\preprint{\vbox{\hbox{SU-4240-634} \hbox{Nov 1996}}}

\maketitle
\begin{abstract}
There are numerous examples of very nearly degenerate states of
opposite parity in molecular physics. The ammonia maser is based on
one such doublet. Theory shows that these parity doublets can occur if
the nuclear shape in the molecule is reflection-asymmetric because the
time scales of the shape and the electronic cloud are
well-separated. Parity doublets occur in nuclear physics as well for
odd $A \sim 219-229$. We discuss the theoretical foundation of these
doublets and on that basis suggest that parity doublets should occur in
particle physics too. In particular they should occur among baryons
composed of ${\it cbu}$ and ${\it cbd}$ quarks.
\end{abstract}
\section{Introduction}
In molecular physics, low energy excitations are rotational bands
stacked on vibrational energies $E_{n}$ (see for example,
\cite{LL-QMNR}). For a molecule with moment of inertia $I$, they have
energies $E_{n} + J(J+1)/2I $ with the angular momentum $J$ assuming
successive values. The separation $E_{n'} - E_{n}$ of vibrational
excitations is much larger than rotational energies. Now if the levels
$(n, J)$ for given $n$ and $J$ are non-degenerate (but for angular
momentum degeneracy), then one of the transitions $(n', J) \rightarrow
(n, J)$ or $(n', J \pm 1) \rightarrow (n, J)$ would be forbidden in
the dipole approximation by parity conservation, and the corresponding
spectral line would be weak. This is so because in this scenario,
states of successive $J$ and same $n$ differ in parity. This is seen,
for example, in the spectrum of $C_2H_2$ \cite{Her2}. 

But there are molecules like $C_2HD$ and $NH_{3}$ where there is no
such intensity alternation, \cite{Her2}. Chemists interpret this
result as an indication that there is a pair of approximately
degenerate levels of opposite parity sitting at each $n$ and
$J$. These parity doublets have also been directly observed for some
molecules like $NH_{3}$ \cite{Her2}, the ammonia maser being based on
just such a doublet \cite{Fey3}.

In nuclear physics, there is evidence for pear-shaped nuclei in the
range odd $A \sim 219-229$ \cite{leashe,leache}. Parity doublets have
been found for these nuclei too \cite{leache,leashe,BohMot2,RinSch}
although their level separation is not small \cite{leache,leashe}.

Parity doublets occur if the shape is reflection-asymmetric. It thus
seems that reflection-asymmetric shapes can lead to approximately
degenerate parity doublets under favorable circumstances.

There is good reason to regard this physical phenomenon as truly
remarkable.  The effective theory of these doublets would be
(approximately) $U(2)$- symmetric even though there is no trace of
such a symmetry in the microscopic Hamiltonian. This $U(2)$
furthermore mixes states of differing parity. So what we have here is
the striking emergence of spontaneous chiral symmetry.  And that is
not all. Below we shall indicate the theory of these doublets (and
elsewhere \cite{balvai} more thoroughly develop it) and point out
their significance for such an apparently remote topic as topology
change in quantum gravity. But our principal concern in this paper is
with a different subject. The above phenomenon has specific
implications for the phenomenology of particle and especially heavy
quark physics, and it is the latter that we focus on in this letter.

Parity doublets occur typically in systems with two differing time
scales. For molecules, the fast variables are electronic and the slow
ones are nuclear. For nuclei, they are the intrinsic and the
rotational degrees of freedom. These systems are amenable to treatment
in the Born-Oppenheimer [B-O] approximation \cite{LL-QMNR}. In this
approximation, there is a simple and vivid manner to understand the
mechanism behind these doublets. Thus, consider for example a molecule
like $C_2HD$ \cite{LL-QMNR,Her2}. It is a linear molecule with $D$ at
one end, and can be approximated by a unit vector $\overrightarrow{n}$
(parallel to the molecule with the tail at $D$) when finding the
rotational levels. The electronic Hamiltonian ${\cal H}_F$ in the B-O
approximation is diagonalized by treating $\overrightarrow{n}$ as
fixed. Now the system as a whole is rotationally invariant, so for
fixed $\overrightarrow{n}$, ${\cal H}_F$ is invariant under rotations
about the axis $\overrightarrow{n}$. If $\overrightarrow{J_F}$ is the
fast variable angular momentum, an eigenstate of ${\cal H}_F$ can be
associated with a definite value of $\overrightarrow{n}.
\overrightarrow{J_F}$. It need not be zero, indeed it will not be so
for an odd number of electrons, as then no component of
$\overrightarrow{J_F}$ has eigenvalue zero. But $\overrightarrow{n}.
\overrightarrow{J_F}$ reverses under parity $\cal P$, so there is
another state with the opposite value of
$\overrightarrow{n}. \overrightarrow{J_F}$ when the latter is
non-vanishing.  When we pass beyond the B-O approximation, the exact
Hamiltonian $\cal H$ mixes these levels, thus creating mutually split
even and odd energy eigenstates.

Now there are of course many shapes in nature. The configuration space
of a shape is just an orbit of the rotation group
\cite{BalBook,balsimwit}. It is thus $SU(2)/H$ for a sub-group $H$ of
$SU(2)$. The molecule is an arrow only if $H=U(1)$. Elsewhere
\cite{balsimwit}, the quantum theory of a generic shape was treated in
detail and it was effectively shown that parity doublets can occur if
the shape lacks reflection symmetry even if $H \neq
U(1)$. [Cf. Section III. The content of that paper is best combined
with \cite{balvai} to reach this conclusion rigorously.] Let us give
an example. If the molecule is a pyramid with the symmetry $Z_{2N}
\subset SU(2)$ around an axis $\overrightarrow{n}$, then the
eigenststates of ${\cal H}_F$ can be associated with definite values
of $\exp[{(2 \pi i \overrightarrow{n}. \overrightarrow{J_F})/N}]$. It
determines helicity $\overrightarrow{n}. \overrightarrow{J_F}$ only
mod $N$.  Under parity, $\exp[{(2 \pi i
\overrightarrow{n}. \overrightarrow{J_F})/N}] \rightarrow \exp[{(-2
\pi i \overrightarrow{n}. \overrightarrow{J_F})/N}]$, and hence there
are parity doublets unless $\exp[{(2 \pi i
\overrightarrow{n}. \overrightarrow{J_F})/N}] = \pm 1$.  Thus an
$N$-fold axis, defining only helicity mod $N$, can also lead to parity
doublets.

Parity doublets are also time-reversal ($\cal T$-) doublets
\cite{balsimwit}. That is because $\cal T$ reverses
$\overrightarrow{J_F}$ and hence $
\overrightarrow{n}. \overrightarrow{J_F}$, just as $\cal P$ does. But
we recall that there could be $\cal T$-doublets both with trivial
parity $+1$, as it happens with staggered conformations
\cite{balsimwit}.

\section{Baryon Physics}

All this could be of concern also to a particle physicist. Thus
tentatively regarding $\it u$ and $\it d$ as light and the remaining
quarks as heavy, the following potential parity-doubled baryon states
come to mind: 1)${\it scu, scd}$ 2)${\it cbu, cbd}$ 3)${\it btu,
btd}$.But there are two important issues to be addressed before we can
entertain the conjecture of parity doublets among these combinations,
namely: 1)the existence of two well- separated time-scales, $T_{slow}$
and $T_{fast}$ associated with the heavy and the light quarks, and  2)the
relative magnitude of $T_{slow}$ and quark lifetimes $\tau$.  Item 1)
is of course the basis of canonical B-O approximation while 2) is
new. It is just that the entire approximation scheme can break down if
a quark decays too fast. It is thus necessary to check that the lifetimes
of quarks are much longer than the dynamical time $T_{slow}$ in the
problem. Below we outline how we treat 1) and 2) and
then summarize the pertinent numbers in tables.

Item1):Assuming that the distance between the two heavy quarks is of the
order of 1 fm, we will estimate $T_{slow}$ as follows. If $I$ is the moment of
inertia of the heavy quark pair, and $J$ its angular momentum, then
$T_{slow} \approx 2 \pi I/J \approx 2 \pi I = 2 \pi \mu R^2$, where
$\mu$ is the reduced mass and $R \approx 1 fm$ is the relative
separation of the heavy quarks. We will estimate $\mu$ and $T_{slow}$
using constituent quark masses, as it is  more appropriate than using 
current quark masses.

As for $T_{fast}$, by the uncertainty principle, the momentum $\it p$
of a fast quark is $\approx 1/R$. It is also $mv/{\sqrt{1-v^2}}$ for a
quark of mass $m$. In this way, we can find $T_{fast} \approx
2R/v$. 

Item 2):Quark lifetime scales as the fifth power of the mass. Crude
estimates for $\tau$ good enough for us can be got by scaling muon
lifetime.  

\begin{figure}
\begin{center}
\begin{tabular}{|c|c|c|}
      \hline
        & Constituent & Constituent    \\   
Quark   & Quark Mass  & Quark Lifetime \\   
        & (GeV)       &   (sec)         \\ 
       \hline
$\it u$ & $\sim$0.3   & $\geq 10^{-6}$  \\
       \hline
$\it d$ & $\sim$0.3   & $\geq 10^{-6}$  \\
       \hline
$\it s$ & $\sim$0.51  & $10^{-6}-10^{-9}$\\ 
       \hline
$\it c$ & 1.1-1.6     &$10^{-11}-10^{-12}$\\
       \hline 
$\it b$ & 4.1-4.5     & $10^{-14}$     \\
       \hline	  
$\it t$ & 170         & $10^{-22}$      \\
       \hline
\end{tabular} \\
\vspace{2mm}
Table 1: Constituent quark masses and their estimated lifetimes.
\end{center}
\end{figure}

\begin{figure}
\begin{center}
\begin{tabular}{|l|c|c|}     
	\hline 
               &                     &                     \\
 Baryon        &  $T_{slow}$         & $T_{slow}/T_{fast}$ \\  
   	       &    (sec)            &        \\             
	\hline
${\it scu/scd}$ & $\sim 10^{-23}$    &        2.9-3.5  \\  
	\hline
${\it sbu/sbd}$ & $\sim 10^{-23}$    &        4.1-4.3  \\  
	\hline
${\it stu/std}$ & $\geq 10^{-22}$    &        4.5      \\  
	\hline
${\it cbu/cbd}$ & $\sim 10^{-23}$    &        8.8-11   \\  
	\hline
${\it ctu/ctd}$ & $\geq 10^{-22}$    &        8.8-17.6 \\  
	\hline
${\it btu/btd}$ & $\geq 10^{-22}$    &        44       \\  
        \hline
\end{tabular} \\
\vspace{2mm}
Table 2: $T_{slow}$ and $T_{slow}/T_{fast}$ for baryons of interest.
\end{center}
\end{figure}

Constituent quark masses and their lifetimes are shown in Table 1, while
numbers for $T_{slow}$ and $T_{slow}/T_{fast}$ are shown in Table 2.    

From the tables, one sees that ${\it cbu}$ and ${\it cbd}$ baryons are
the best candidates to search for parity doublets. Since $J_F$, the total 
angular momentum of the light quark, is necessarily half-integral, we expect 
that parity doublets will occur. In addition to
parity doublets, the model of course predicts normal rotational
excitations. Their splitting would be of the order of $1/2I \simeq$
100 MeV and can be looked for experimentally. It is difficult to
estimate the energy difference between the parity doublets. It could
be of the order of 100 MeV (that is,of the order of rotational
excitation energies as in nuclear physics) or smaller. If these levels
are split by more than the pion mass, they can be detected by s-wave
pion decay (or some other strong decay) of the higher state. If the
mass difference is not so much, and the spin is 1/2, then the dominant
decay will involve the emission of photons via a pseudotensor
coupling. However these observations may not give the best signals for
the detection of parity doublets. In fact, we can find none, comparable
in elegance to study of intensity alternation patterns in molecular
physics alluded to previously, for the detection of such doublets in
particle physics.

In the B-O approximation, the heavy quarks are not in a definite
orbital angular momentum state, in contrast to what is found in quark
models. For this and for other reasons, the relation of the B-O
and quark model states is intricate and will be elaborated in \cite{balvai}.

\section{Heavy Mesons, Skyrmions and Monopoles}

Baryons are not the only favorable systems for parity
doublets. Literature abounds in speculation \cite{tor,manwis}
suggesting the existence of heavy meson bound states. They can involve
distinct heavy mesons too. These can be the slow variables and
suitable excitations (like the $\rho$ or the $\omega$ meson) can be the 
fast ones, and we may have parity doublets again.

These doublets may also appear in the physics of Skyrmions and
monopoles. For the former, there now exist elaborate simulations of
static configurations for differing baryon numbers
\cite{bratowcar,car}. They are found to occur as regular solids with
discrete symmetry groups. We can also imagine that further
calculations will show static configurations such as a pear, with
$U(1)$ symmetry group. Excitations with spin, like a $\rho$ or an
$\omega$, or even a nucleon,which can have non-zero helicity
$\overrightarrow{n}.  \overrightarrow{J_F}$,could then lead to parity
doublets.

Of equal interest to the above Skyrmion configurations are the static
monopole configurations with symmetries under rotation subgroups
\cite{hitmanmur,suthou1,suthou2}.  They can occur in grand unified
models. By attaching fast constituents such as a spin 1/2 quark, we
can hope to create parity doublets in these systems, just as in
molecular physics.

\section{A Remark and a Reminder}

Effects of heavy particle (slow core) spins are neglected in the B-O
approximation. They could lead to additional degeneracies and may
require future consideration.

It is crucial in these considerations that the slow configuration is
reflection-asymmetric for parity doublets to occur. They would not
occur in ${\it ccu}$, as ${\it c-c}$ is described as a headless
arrow. They would also not occur for staggered conformations which are
reflection-symmetric even though they can have a doublet structure
mixed by $\cal T$. It would be most striking to encounter these $\cal T$-
doublets, predicted naturally theoretically, in chemistry, nuclear and
particle physics.

\section{Final remarks}

It is appropriate to conclude by outlining certain more formal
considerations which we shall elsewhere study in greater depth
\cite{balvai}.

The quantum theory of three-dimensional shapes, that is, quantization
on configuration spaces $Q=SU(2)/H$ was studied in \cite{balsimwit}.
As is well-known \cite{BalBook,balsimwit}, it is not unique, there
being a distinct quantization for each unitary irreducible
representation (UIR) $\rho$ of $H$. For a particular $\rho$, the
domain of the shape Hamiltonian ${\cal H}_S$ consists of sections of a
vector bundle associated with $\rho$. These domains and hence the
corresponding quantum theories are different for different $\rho$.
Now, it so happens for reflection-asymmetric shapes that $\cal P$ can
map $\rho$ to an inequivalent UIR ${\rho}_{\cal P}$ and so the quantum
theory to an inequivalent one. Quantum theory thereby spoils classical
$\cal P$- invariance, in precisely the same manner that the presence of the 
topological $\theta$-term in  QCD (for $\theta \neq 0, \pi$) breaks it 
\cite{caldasgro,jacreb}. Also ${\rho}_{\cal P} = {\rho}^*$ \cite{balsimwit}, 
so $\cal T$ is violated by quantization, but not $\cal{PT}$. But the strange 
behavior of staggered conformations noted earlier is unlike anything we know 
of in conventional particle theory.

These results on shapes are paradoxical. There is no $\cal P$- or
$\cal T$- violation in molecular physics while shapes(slow cores) with
$\cal P$- or $\cal T$- violating $\rho$ do occur in nature. How then
is this paradox resolved?

The resolution is as follows. Let us at the start assume that the
domain $V^{({\rho}_0)}$ of the total Hamiltonian ${\cal H}={\cal H}_S
+{\cal H}_F$ is associated with the trivial representation ${\rho}_0$
that harms neither $\cal P$ nor $\cal T$. The domain of ${\cal H}_S$
is then also the domain associated with ${\rho}_0$. An eigenstate
${\psi}^{({\overline{\rho}})}_F$ of the fast Hamiltonian ${\cal H}_F$
is the section of a vector bundle over $Q$ in the B-O
approximation~\footnote{The Berry phase shows this
result.}~\cite{ShaWil,mooshawil} (the superscripts on wave functions
will indicate the UIR) and it can happen that this bundle is twisted
and is associated with a UIR ${\overline{\rho}}$. The B-O slow
Hamiltonian is not ${\cal H}_S$, it must be obtained from averaging ${\cal H}$
over ${\psi}^{({\overline{\rho}})}_F$, and when that is done, the
emergent slow Hamiltonian ${\hat{\cal H}}_S$ contains a connection and
has a domain associated with the UIR $\rho$, which is the complex
conjugate of $\overline{\rho}$. (A result along these lines is in
\cite{ShaWil,mooshawil}). So an eigenstate ${\psi}^{({\rho})}_S$ of
${\hat{\cal H}}_S$ corresponds to $\rho$ and the product
wave function ${\psi}={\psi}^{({\rho})}_S 
{\psi}^{({\overline{\rho}})}_F$ corresponds to $\rho \otimes
\overline{\rho}$.  But $\cal H$ and ${\cal H}_S$ act on the total
wave function and their domain can only correspond to ${\rho}_0$. That
is now easily arranged as ${\rho}_0$ occurs in the reduction of $\rho
\otimes \overline{\rho}$. The correct total wave function in the B-O
approximation is thus the projection ${\chi}^{({\rho}_0)} = {\bf
P}[{\psi}^{({\rho})}_S {\psi}^{({\overline{\rho}})}_F]$ of
$\psi$ to $V^{({\rho}_0)}$.  If ${\rho}_{\cal P} = \overline{\rho}$,
the parity transform ${\cal P} {\chi}^{({\rho}_0)}$ of
${\chi}^{({\rho}_0)}$ is of the form ${\bf
P}[{\psi}^{({\overline{\rho}})}_S {\psi}^{({\rho})}_F] \in
V^{({\rho}_0)}$. It is still in the domain of $\cal H$ and ${\cal
H}_S$, so there is no question of $\cal P$-violation. The same goes
for $\cal T$. The doublets with definite $\cal P$ in the leading
approximation are linear combinations of ${\chi}^{({\rho}_0)}$ and
${\cal P}{\chi}^{({\rho}_0)}$.

A remarkable feature of the B-O approximation, occasionally
appreciated before, is that eigenstates of ${\hat{\cal H}}_S$ may be
states with helicity \cite{mooshawil}, even spinorial states,even
though those of ${\cal H}_S$ are tensorial zero-helicity ones. This
happens if for example, the configuration space $Q$ is the two-dimensional 
sphere $S^2$ = $\{ \overrightarrow{n} \}$. If the
helicity $\overrightarrow{n}. \overrightarrow{J_F} = -K$ of the ${\cal
H}_F$ eigenstate is non-vanishing,then the slow wave function is a section 
of the monopole bundle with helicity (Chern number) $K$. The slow
wave function is then spinoral if $K \in (2Z+1)/2$. A spinorial slow
eigenfunction can get converted to a tensorial one too under suitable
conditions.

Now suppose that the fast variables (with UIR $\overline{\rho}$)
cannot be seen by current experiments, perhaps because their
excitations are too energetic. They can still leave a trace in the
slow system by twisting its bundle from ${\rho}_0$ to $\rho$ or
changing its prior twist, and perhaps even altering its tensorial or
spinorial character. If without our being aware, the fast variable for
UIR $\overline{\rho}$ is replaced by another for UIR ${\rho}'$, the
slow bundle too is changed thereby. This is an effective topology
change, but at a quantum level, for the Hamiltonian ${\cal H}_S$. The
topology change of classical configuration space, frantically sought
in gravity, cannot be achieved in this manner. That would require
another mechanism like cobordism in functional integrals \cite{dowsor}
or domain changes (of a new sort) of the Hamiltonian
\cite{balbimmarsim}.

M. V. N. Murthy has been exceptionally helpful to us in the course of
this work, while Charlie Nash pointed out to us that the ammonia maser
is based on a parity doublet. We are sincerely grateful to them, and
also to Brian Dolan, Carl Rosenzweig and one of the referees of Physical 
Review Letters for important comments. Charilaos Anezeris, Kumar Gupta and 
Al Stern participated in the early stages of this work. We have benefited 
from this collaboration and also from conversations with our experimental 
group, especially Marina Artuso, Nahmin Horowitz, Giancarlo Moneti and
Sheldon Stone. This work was supported in part by the US DOE under
contract number DE-FG02-85ER40231.

\bibliography{quark1}
\bibliographystyle{prsty}

\end{document}